\documentclass[aps,prb,twocolumn,showpacs,groupedaddress, amsmath]{revtex4}

\usepackage{graphicx}
\usepackage{amssymb}

\begin{document}


\title{Aging dip and cumulative aging in hierarchical successive transition of stage-2 CoCl$_{2}$ graphite intercalation compound}


\author{Masatsugu Suzuki}
\email[]{suzuki@binghamton.edu}
\affiliation{Department of Physics, State University of New York at Binghamton, Binghamton, New York 13902-6000}

\author{Itsuko S. Suzuki}
\email[]{itsuko@binghamton.edu}
\affiliation{Department of Physics, State University of New York at Binghamton, Binghamton, New York 13902-6000}


\date{\today}

\begin{abstract}
Memory and aging behaviors in stage-2 CoCl$_{2}$ GIC ($T_{cu}$ = 8.9 K and $T_{cl}$ = 6.9 K) have been studied using low frequency ($f$ = 0.1 Hz) AC magnetic susceptibility ($\chi^{\prime}$ and $\chi^{\prime\prime}$) as well as thermoremnant magnetization. There occurs a crossover from a cumulative aging (mainly) with a partial memory effect for the domain-growth in the intermediate state between $T_{cu}$ and $T_{cl}$ to an aging and memory in a spin glass phase below $T_{cl}$. When the system is aged at single or multiple stop and wait processes at stop temperatures $T_{s}$'s below $T_{cl}$ for wait times $t_{s}$, the AC magnetic susceptibility shows single or multiple aging dips at $T_{s}$'s on reheating. The depth of the aging dips is logarithmically proportional to the wait time $t_{s}$. Very weak aging dip between $T_{cu}$ and $T_{cl}$ indicates the existence of a partial memory effect. The sign of the difference between the reference cooling and reference heating curves changes from positive to negative on crossing $T_{cl}$ from the high $T$-side. The time dependence of $\chi^{\prime}$ and $\chi^{\prime\prime}$ below $T_{cl}$ is described by a scaling function of $\omega t$. its a local maximum at 6.5 K just below $T_{cl}$, and drastically decreases with increasing $T$. The nature of the cumulative aging between $T_{cl}$ and $T_{cu}$ is also examined.
\end{abstract}


\maketitle



\section{\label{intro}INTRODUCTION}
Memory and aging behaviors have been observed in many kinds of glassy materaials such as spin glasses (SG's),\cite{ref01,ref02,ref03,ref04,ref05,ref06,ref07,ref08,ref09,ref10} reentrant ferromagnets,\cite{ref11,ref12,ref13} ferroelectrics,\cite{ref14,ref15,ref16} polymer glasses,\cite{ref17} and ferromagnetic nanoparticles.\cite{ref18} This is mainly due to a consequence of disorder and frustration of systems. The memory effect seems to be a feature which is universal to the aging behaviors in glassy materials, whereas the hysteresis is present in relaxor ferroelectrics and polymer glasses, but not in all kinds of SG's. Comparative studies can shed light on what is universal in the memory effect of the disordered systems and provide some ideas of theoretical interpretation. 

It is well known that the aging dynamics in SG systems strongly depend on the history of the system after quenching from a temperature well above the spin freezing temperature $T_{SG}$.\cite{ref01,ref02,ref03,ref04,ref05,ref06,ref07,ref08,ref09,ref10} In spite of the fact that the aging history is reflected by the underlying nature of each disordered system, there are some essential features which seem to be common to the memory effect of the aging behavior in SG systems. First, the system is quenched from a temperature well above $T_{SG}$ and is isothermally aged at $T_{s}$ for a wait time $t_{s}$. The age of the system at $T_{s}$ is equal to $t_{s}$. Suppose that the temperature $T$ is set back to $T_{s}$ after it is temporarily changed to $T_{0}$. When $T_{0}$ is lower than $T_{s}$, the age of the system at $T=T_{s}$ remains unchanged (the memory effect). When $T_{0}$ is higher than $T_{s}$, the age of the system at $T=T_{s}$ becomes reset to zero (the rejuvenation effect). 

Experimentally, the AC magnetic susceptibility (the dispersion $\chi^{\prime}$ and the absorption $\chi^{\prime\prime}$) can be used to determine the nature of the aging dynamics in SG's.\cite{ref01,ref02,ref03,ref04,ref05} The system ages as they approach thermal equilibrium and settle into lower energy states. This aging is observed as a decrease in both $\chi^{\prime}$ and $\chi^{\prime\prime}$ when the system is isothermally aged at a stop temperature $T_{s}$. Aging a spin glass at $T_{s}$ reduces $\chi^{\prime}$ and $\chi^{\prime\prime}$ only in the immediate vicinity of $T_{s}$, creating a so-called aging dip. So long as $T$ is kept below $T_{s}$, there will be memory of this aging dip. In addition, SG's also approximately exhibits $\omega t$ scaling, where aging of $\chi^{\prime}$ and $\chi^{\prime\prime}$ is a function of the product $\omega t$ and not $\omega$. The aging and the response come from the same sort of degrees of freedom, as in hierarchical kinetic schemes. 

In our previous two papers,\cite{ref19,ref20} we have undertaken an extensive study on a series of the time dependence of the zero-field cooled (ZFC) magnetization $M_{ZFC}$ of stage-2 CoCl$_{2}$ graphite intercalation compound (GIC), after the ZFC aging protocol. This compound undergoes hierarchical successive transitions at $T_{cu}$ (= 8.9 K) and $T_{cl}$ (= 7.0 K). We find that the time dependence of the relaxation rate $S_{ZFC}(t)$ = $(1/H)$d$M_{ZFC}(t)$/d$\ln t$ below $T_{cl}$ exhibits a single peak at a characteristic time close to a wait time $t_{w}$, as is typically observed in the aging dynamics of a SG. In contrast, $S_{ZFC}(t)$ has two distinguished peaks at different characteristic times between $T_{cu}$ and $T_{cl}$, indicating the coexistence of two correlated regions. One goes into a SG-like ordered state at $T_{cu}$. The other goes into a similar glassy state below $T_{cu}$, but of much shorter relaxation time. The intermediate state is a ferromangetic-like glassy phase characterized by an intra-cluster (ferromagnetic) order and an inter-cluster disorder.\cite{ref20} 

Our purpose in the present work is to study (i) the nature of aging dip for $T_{s}<T_{cl}$, (ii) a cumulaticve aging and a partial memory effect for $T_{cl}<T_{s}<T_{cu}$, and (iii) a hysteresis behaviors, in the stage-2 CoCl$_{2}$ GIC. The $T$ dependence of the AC magnetic susceptibility is measured with increasing $T$ after a single stop and wait (SSW) process. The stop temperature $T_{s}$ and the stop time $t_{s}$ are changed as parameters. Our results will be compared with those of reentrant spin glasses (RSG's)\cite{ref11,ref12,ref13} and relaxor ferroelectrics.\cite{ref16} 

\section{\label{exp}EXPERIMENTAL PROCEDURE}
We used a sample of stage-2 CoCl$_{2}$ GIC based on a single crystal kish graphite which was used in previous papers.\cite{ref19,ref20} The AC magnetic susceptibility and DC magnetization were measured using a SQUID magnetometer (Quantum Design, MPMS XL-5) with an ultra low field capability option. The time and temperature dependence of the AC magnetic susceptibility was measured after appropriate zero-field cooling protocols. Typically we used an AC frequency ($f$ = 0.1 Hz) and the amplitude of the AC field ($h$ = 0.1 Oe). The temperature dependence of the thermoremnant (TRM) magnetization was also measured after appropriate FC protocol. The detail of the experimental procedure of each measurement will be described in Sec. \ref{result} and the figure captions. 

\section{\label{result}RESULT}
\subsection{\label{resultA}$M_{TRM}$ for the DSW and SSW processes}
The TRM magnetization in stage-2 CoCl$_{2}$ GIC was measured with increasing $T$ after the following FC protocol with a double stop and wait process (DSW). First the system was aged at 50 K for $1.2\times 10^{3}$ sec in the presence of a magnetic field $H_{c}$ (= 1 Oe). The cooling of the system was resumed in the presence of the same field $H_{c}$ from 50 K to the first stop temperature $T_{s1}$. It was again aged at $T_{s1}$ for a stop time $t_{s1}$ ($= 3.0\times 10^{4}$ sec). The cooling was resumed from $T_{s1}$ to the second stop temperature $T_{s2}$ ($<T_{s1}$). The system was aged at $T_{s2}$ for a stop time $t_{s2}$ ($= t_{s1}$). The cooling of the system was again resumed to 2 K. Immediately after the magnetic field is turned off, the TRM magnetization was measured with increasing $T$ from 2 to 15 K. The $T$ dependence of the magnetization thus obtained is denoted by $M_{TRM}(T_{s1},t_{s1};T_{s2},t_{s2},T)$. The $T$ dependence of $M_{TRM}^{ref}(T)$ as a reference curve was also measured with increasing $T$ after the FC protocol ($H_{c}$ = 1 Oe) without any stop and wait process. The genuine TRM magnetization is defined as 
\begin{eqnarray}
&\Delta& M_{TRM}(T_{s1},t_{s1};T_{s2},t_{s2},T) \nonumber\\
&=& M_{TRM}(T_{s1},t_{s1};T_{s2},t_{s2},T) - M_{TRM}^{ref}(T).
\label{eq01}
\end{eqnarray}

\begin{figure}
\includegraphics[width=6.5cm]{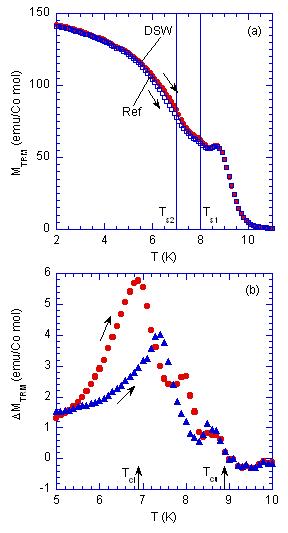}
\caption{\label{fig01}(Color online) (a) $M_{TRM}$ vs $T$, which was measured on reheating after double-stop and wait (DSW) processes during the FC protocol, and on heating after the FC protocol without stop and wait process. (i) The reheating curve ($M_{TRM}$) denoted by closed squares. The system was cooled from 50 to 8.0 K in the presence of $H_{c}$ (= 1 Oe). It was aged at a stop temperature $T_{s1}$ = 8.0 K for a wait time $t_{s1} = 3.0\times 10^{4}$ sec. Then the cooling was resumed. It was again isothermally aged at $T_{s2}$ = 7.0 K for $t_{s2} = 3.0\times 10^{4}$ sec. Then it was cooled again down to 2.0 K. After the magnetic field is turned off, the TRM magnetization was measured with increasing $T$. (ii) Reference heating curve ($M_{TRM}$) denoted by open squares. The system was cooled in the presence of $H_{c}$ (= 1 Oe) from 50 to 2.0 K without any stop and wait process. After the magnetic field is turned off, the TRM magnetization was measured with increasing $T$. (b) $T$ dependence of $\Delta M_{TRM}$ (defined by Eq.(\ref{eq01})) obtained from the DSW experiments. $H_{c}$ = 1.0 Oe. The DSW process at $T_{s1}$ = 8.0 K ($t_{s1} = 3.0\times 10^{4}$ sec) and $T_{s2}$ = 7.0 K ($t_{s2} = 3.0\times 10^{4}$ sec) (closed circles), the DSW process at $T_{s1}$ = 8.5 K ($t_{s1} = 3.0\times 10^{4}$ sec) and $T_{s2}$ = 7.5 K ($t_{s2} = 3.0\times 10^{4}$ sec) (closed triangles).}
\end{figure}

Here we present our results on the $T$ dependence of the TRM magnetization when the FC protocol is interrupted by the DSW process. Two kinds of the DSW process are used: (i) $T_{s1}$ = 8.0 K and $T_{s2}$ = 7.0 K, (ii) $T_{s1}$ = 8.5 K and $T_{s2}$ = 7.5 K for $t_{s1}=t_{s2}=t_{s} = 3.0\times 10^{4}$ sec. Figure \ref{fig01}(a) shows the $T$ dependence of the TRM magnetization $M_{TRM}(T_{s1},t_{s1};T_{s2},t_{s2},T)$ with the DSW process [the case (i)] and the reference curve $M_{TRM}^{ref}(T)$ without stop and wait process. Figure \ref{fig01}(b) shows the plot of $\Delta M_{TRM}(T_{s1},t_{s1};T_{s2},t_{s2},T)$ as a function of $T$ for the DSW process for both cases (i) and (ii). We find that there appear two peaks at $T_{s1}$ and $T_{s2}$. This implies that the ages of the system memorized at $T_{s1}$ and $T_{s2}$ becomes frozen in on further cooling from $T_{s2}$ to 2.0 K and are retrieved at $T_{s2}$ and T$_{s1}$ on heating from 2.0 K to 15 K. The system remembers its age or shows a memory phenomenon. The height of aging peak decreases with increasing $T$ for $T_{cl}\le T\le T_{cu}$ and reduces to zero at $T_{cu}$, giving an evidence for the occurrence of partial memory effect.

The following rules may be derived from this experiment. (i) The age of the system recorded at $T_{s1}$ is independent of that recorded at $T_{s2}$ (the temperature independence of the aging effect). (ii) The age of the system at $T_{s1}$ is preserved for $T<T_{s1}$ and the age of the system at $T_{s2}$ is preserved for $T<T_{s2}$ (the memory effect). (iii) The age of the system at $T_{s2}$ is rejuvenated for $T>T_{s2}$ and the age of the system at $T_{s1}$ is rejuvenated for $T>T_{s1}$ (the rejuvenation effect). Similar behavior has been observed by Mathieu et al.\cite{ref06} on the 3D Ising SG Fe$_{0.55}$Mn$_{0.45}$TiO$_{3}$. 

\subsection{\label{resultB}$T$ dependence of $\chi^{\prime}$ and $\chi^{\prime\prime}$ during the SSW process}

\begin{figure}
\includegraphics[width=7.0cm]{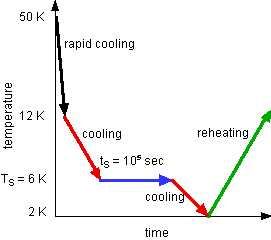}
\caption{\label{fig02}(Color online) Schematic diagram of the experimental procedure for the SSW memory experiment in the AC magnetic susceptibility.}
\end{figure}

\begin{figure}
\includegraphics[width=7.0cm]{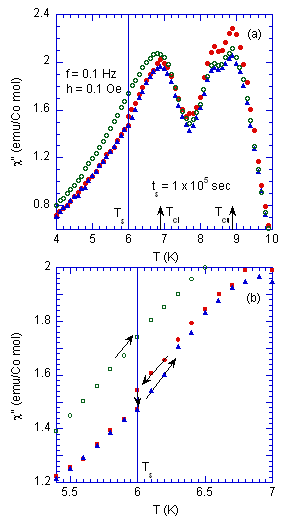}
\caption{\label{fig03}(Color online) (a) and (b) $T$ dependence of $\chi^{\prime\prime}$. $f$ = 0.1 Hz. $h$ = 0.1 Oe. $H$ = 0. (i) The cooling curve (closed circles). After the ZFC protocol from 50 to 12 K at $H$ = 0, $\chi^{\prime\prime}$ was measured with decreasing $T$ from 12 to 6.0 K at $H$ = 0. The system was isothermally aged at $T_{s}$ = 6.0 K for $t_{s} = 1.0\times 10^{5}$ sec (SSW process). The measurement of $\chi^{\prime\prime}$ was resumed with decreasing $T$ from 6.0 to 2.0 K. (ii) The reheating curve (closed triangles). Subsequent measurement of $\chi^{\prime\prime}$ was made with increasing $T$ from 2.0 to 12.0 K at $H$ = 0. (iii) The reference heating curve (open circles). After the ZFC protocol (quenching from 50 to 2 K) at $H$ = 0, $\chi^{\prime\prime}_{refh} $ was measured with increasing $T$ from 2.0 to 12.0 K at $H$ = 0. Note that the reference cooling curve ($\chi_{refc}^{\prime}$ and $\chi_{refc}^{\prime\prime}$) above 6.0 K is the same as the cooling curve with the SSW process at $T_{s}=6.0$ K.}
\end{figure}

Figure \ref{fig02} shows a schematic diagram of the experimental procedure for the SSW memory experiment in the AC magnetic susceptibility. The history of the system was reinitialized by heating the sample at 50 K well above $T_{cu}$. Immediately after the system was annealed at 50 K for $1.2\times 10^{3}$ sec, it was rapidly cooled from 50 to 12 K at $H$ = 0 (the ZFC protocol). The measurement of the dispersion $\chi^{\prime}$ and absorption $\chi^{\prime\prime}$ was carried out with decreasing $T$ from 12 to 6.0 K at $H$ = 0 by a step temperature $\Delta T$ = 0.1 K. The system was isothermally aged at $T_{s}$ = 6.0 K for $t_{s}$ = $1.0\times 10^{5}$ sec (SSW process). The measurement of $\chi^{\prime}$ and $\chi^{\prime\prime}$ was resumed with decreasing $T$ from 6.0 to 2.0 K (the cooling curve). Immediately after the lowest temperature (2.0 K) was reached, the measurement of $\chi^{\prime}$ and $\chi^{\prime\prime}$ was carried out with increasing $T$ from 2.0 to 12.0 K at $H$ = 0 (the reheating curve). For comparison, the reference heating curve ($\chi^{\prime}_{refh} $ and $\chi^{\prime\prime}_{refh}$) was also measured as follows. After the ZFC protocol (annealing at 50 K for $1.2\times 10^{3}$ sec and quenching from 50 to 2 K at $H$ = 0), the measurement of $\chi^{\prime}_{refh}$ and $\chi^{\prime\prime}_{refh}$ was carried out with increasing $T$ from 2.0 to 12.0 K at $H$ = 0. Note that the cooling curve with the stop and wait process at 6.0 K for $6.0\le T\le 12.0$ K is used as the reference cooling curve ($\chi^{\prime}_{refc}$ and $\chi^{\prime\prime}_{refc}$), since the cooling curve above $T_{s}$ is not affected by the SSW process at $T_{s}$. 

\begin{figure*}
\includegraphics[width=12.0cm]{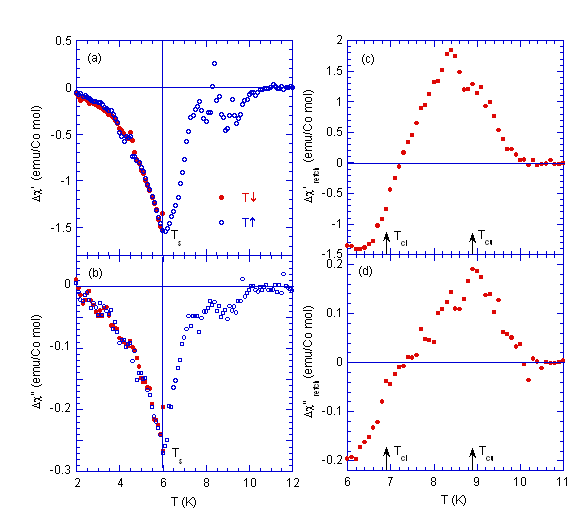}
\caption{\label{fig04}(Color online) (a) and (b) $T$ dependence of $\Delta\chi^{\prime}$ ($= \chi^{\prime}-\chi^{\prime}_{refh}$) and $\Delta\chi^{\prime\prime}$ ($= \chi^{\prime\prime}-\chi^{\prime\prime}_{refh}$) which is obtained as a subtraction of the reheating curves from the reference heating curves ($\chi^{\prime}_{refh}$ and $\chi^{\prime\prime}_{refh}$). $T_{s}=6.0$ K. $t_{s}=1.0\times 10^{5}$ sec. See Figs.~\ref{fig03}(a) and (b). The reheating curves of $\chi^{\prime}$ and $\chi^{\prime\prime}$ (closed circles) were measured from 2.0 to 12.0 K subsequently after the measurement of the cooling curve. $f$ = 0.1 Hz and $h$ = 0.1 Oe. $H$ = 0. (c) and (d) $T$ dependence of $\Delta\chi^{\prime}_{refch}=\chi^{\prime}_{refc}-\chi^{\prime}_{refh}$ and $\Delta\chi^{\prime\prime}_{refch}=\chi^{\prime\prime}_{refc}-\chi^{\prime\prime}_{refh}$, where the reference cooling curves $\chi^{\prime}_{refc}$ and $\chi^{\prime\prime}_{refc}$ were measured with decreasing $T$ from 12.0 to 6.0 K after the ZFC protocol from 50 to 12.0 K. $f$ = 0.1 Hz. $h$ = 0.1 Oe.}
\end{figure*}

Figure \ref{fig03}(a) and (b) show the $T$ dependence of $\chi^{\prime\prime}$, where $f$ = 0.1 Hz and $h$ = 0.1 Oe. which consists of the cooling curve ($T = 12.0 \rightarrow$ 2.0 K) with the SSW process at $T_{s}$ = 6.0 K for $t_{s} = 1.0\times 10^{5}$ sec, the reheating curve ($T = 2.0 \rightarrow 12.0$ K), and the reference heating curve ($T = 2.0 \rightarrow 12.0$ K). Except at the aging of the system at $T_{s}$ as a part of the SSW process, the average cooling and heating rates are constant (1.3 K/hour). The reheating curve lies significantly below the reference heating curve below $T_{cl}$. The cooling curve at $T_{s}$ = 6.0 K relaxes downward due to the aging for $t_{s} = 1.0 \times 10^{5}$ sec. Figures \ref{fig04}(a) and (b) show the $T$ dependence of $\Delta \chi^{\prime}$ ($=\chi^{\prime}-\chi^{\prime}_{refh}$) and $\Delta \chi^{\prime\prime}$ ($=\chi^{\prime\prime}-\chi^{\prime\prime}_{refh}$) for the reheating curve ($T = 2 \rightarrow 12$ K) and for the cooling curve ($T = 6.0 \rightarrow 2.0$ K) defined only below $T_{s}$. These figures indicate how these curves are affected by the SSW process at $T_{s}$ = 6.0 K. We find that both $\Delta \chi^{\prime}$ and $\Delta \chi^{\prime\prime}$ take a local minimum (an aging dip) at $T$ = $T_{s}$. The shape of the aging dip in $\Delta \chi^{\prime\prime}$ is nearly symmetric with respect to the axis of $T$ = $T_{s}$ = 6.0 K. The appearance of the aging dip at $T$ = $T_{s}$ indicates that the memory imprinted during the ZFC protocol with the SSW process at $T_{s}$ is revealed on reheating. Note that the reheating curve coincides with the cooling curve below $T_{s}$. 

\subsection{\label{resultC}Hysteresis of reference cooling and reference heating curves in $\chi^{\prime}$ and $\chi^{\prime\prime}$}
In Fig.\ref{fig03}, the reference cooling curve without any stop and wait process was not measured over the temperature range between 12.0 and 2.0 K. However, we note that the cooling curve with the stop and wait process at $T_{s}$ = 6.0 K for $t_{s} = 1.0\times 10^{5}$ sec coincides with the reference cooling curve at least above 6.0 K. Figures \ref{fig04}(b) and (c) shows the $T$ dependence of the difference $\Delta \chi^{\prime}_{refch}$ and $\Delta \chi^{\prime\prime}_{refch}$ for $T>T_{s}$ = 6.0 K which are the difference between the reference cooling and reference heating curves, 
\begin{equation}
\Delta \chi^{\prime}_{refch} =\chi^{\prime}_{refc}-\chi^{\prime}_{refh}
\text{ and }
\Delta \chi^{\prime\prime}_{refch} =\chi^{\prime\prime}_{refc} -\chi^{\prime\prime}_{refh},
\label{eq02}
\end{equation}
respectively. These curves are not affected by the SSW process at $T=T_{s}$. We find that the $T$ dependence of $\Delta \chi^{\prime}_{refch} $ is similar to that of $\Delta \chi^{\prime\prime}_{refch}$, although the magnitudes are rather different. Both $\Delta\chi^{\prime}_{refch}$ and $\Delta\chi^{\prime\prime}_{refch}$ are negative below 7.2 K ($\gtrsim T_{cl}$), showing positive local maxima around 8.4 and 8.9 K (= $T_{cu}$), and reduces to zero above 10.1 K. The positive sign of $\Delta\chi^{\prime}_{refch}$ between $T_{cu}$ and $T_{cl}$ suggests that the reference heating state is more stable than the reference cooling state. The negative sign of $\Delta \chi^{\prime}_{refch} $ below $T_{cl}$ indicates that the cooling reference state is more stable than the reference heating state. We note that the reference heating curve ($\chi^{\prime\prime}_{refh} $ vs $T$) lies significantly above the reference cooling curve ($\chi^{\prime\prime}_{refc}$ vs $T$ with $f$ = 1 Hz and $h$ = 0.1 Oe) for Ag (11 at \% Mn, Jonsson et al.\cite{ref02}) (except close to the lowest temperatures), which is indicative of $\Delta\chi^{\prime\prime}_{refch}<0$ below $T_{SG}$. The change of sign in $\Delta \chi^{\prime\prime}_{refch}$ is observed in reentrant ferromagnet CdCr$_{1.8}$In$_{0.2}$S$_{4}$ ($T_{c}$ = 50 K, $T_{RSG}$ = 18 K) by Dupuis et al.,\cite{ref12} when the reference cooling and reference heating curves are recorded above the starting temperature $T_{0}$ = 20 K. They have found that $\Delta \chi _{refch} $ is negative for $T_{0}<T<35$ K and that $\Delta\chi_{refch}$ is positive for $35<T<50$ K. 

We also note that in stage-2 CoCl$_{2}$ GIC, the $T$ dependence of $\Delta \chi^{\prime}_{refch}$ shown in Fig.~\ref{fig04}(c) is very similar to that of the nonlinear AC magnetic susceptibility $\chi_{3}^{\prime}$ of the ($h$ = 0.8 Oe and $f$ = 11 Hz), which was reported by Matsuura and Hagiwara,\cite{ref21} and Miyoshi et al.\cite{ref22} The characteristic feature of $\chi_{3}^{\prime}$ vs $T$ curve indicates that the total magnetic symmetry is not broken at $T_{cu}$ but that it is broken at $T_{cl}$. It is theoretically predicted that $\chi_{3}^{\prime}$ is positive for the FM phase, while it is negative for the SG phase.\cite{ref23,ref24} 

\subsection{\label{resultD}Scaling of the relaxation of $\chi^{\prime}(\omega ,t)$ and $\chi^{\prime\prime}(\omega ,t)$}

\begin{figure}
\includegraphics[width=7.0cm]{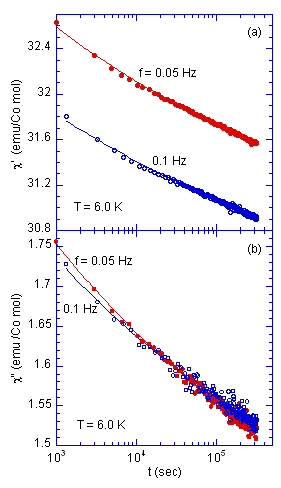}
\caption{\label{fig05}(Color online) Relaxation of (a) $\chi^{\prime}(\omega ,t)$ and (b) $\chi^{\prime\prime}(\omega ,t)$ with time $t$ at $T$ = 6.0 K after the ZFC protocol from 50 to 6.0 K. $t$ = 0 is a time after $T$ becomes stable at 6.0 K. $h$ = 0.1 Oe. $H$ = 0. $f$ = 0.05 and 0.1 Hz. The solid lines denote least-squares fitting curves of the data to the power-law form given by Eqs.(\ref{eq03}) and (\ref{eq04}).}
\end{figure}

\begin{figure}
\includegraphics[width=7.0cm]{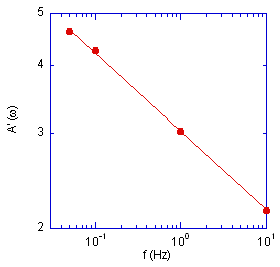}
\caption{\label{fig06}(Color online) $f$ dependence of the amplitude $A^{\prime}(\omega )$ (listed in Table \ref{Table01}) in $\chi^{\prime}(\omega ,t)$ vs $t$. $T$ = 6.0 K. $\omega = 2\pi f$. The amplitude $A^{\prime}(\omega )$ is defined by Eq.(\ref{eq03}). The solid line denotes a least-sqares fitting curve of the data to Eq.(\ref{eq05}) with $\mu^{\prime}=0.144\pm 0.003$.}
\end{figure}

\begin{table}
\caption{\label{Table01}Frequency dependence of the parameters $b^{\prime}$, $\chi _{0}^{\prime}(\omega )$, $A^{\prime}(\omega )$. The change of $\chi^{\prime}(\omega ,t)$ with $t$ due to aging is defined by Eq.(\ref{eq03}).}
\begin{ruledtabular}
\begin{tabular}{llll}
$f$ (Hz) & $b^{\prime}$ & $\chi_{0}^{\prime}(\omega )$ (emu/Co mol)& $A^{\prime}(\omega )$ \\
0.05 & $0.101 \pm 0.004$ & $30.28 \pm 0.07$ & $4.635 \pm 0.023$\\ 
0.1 & $0.079 \pm 0.008$ & $29.34 \pm 0.17$ & $4.266 \pm 0.069$\\ 
1 & $0.090 \pm 0.006$ & $27.75 \pm 0.08$ & $3.022 \pm 0.017$\\ 
10 & $0.102 \pm 0.028$ & $25.91 \pm 0.20$ & $2.16 \pm 0.07$\\
\end{tabular}
\end{ruledtabular}
\end{table}

We measured the $t$ dependence of $\chi^{\prime}(\omega ,t)$ and $\chi^{\prime\prime}(\omega ,t)$ at $T=T_{s}$ (= 6.0 K) at various frequency $f$ (= 0.05 -- 50 Hz) for $0\le t\le 2\times 10^{5}$ sec, where $\omega$ is the angular frequency ($\omega = 2\pi f$), $h$ = 0.1 Oe, and $H$ = 0 Oe. The signal becomes noisy at sufficiently long times for frequencies higher than 10 Hz. The system was annealed at 50 K for $1.2\times 10^{3}$ sec as an initialization. Immediately after the system was rapidly cooled from 50 to 6.0 K (the ZFC protocol), $\chi^{\prime}(\omega ,t)$ and $\chi^{\prime\prime}(\omega ,t)$ were measured simultaneously as a function of the time $t$, where $t$ = 0 is a time when the temperature of the system becomes stable at 6 K. Figures \ref{fig05}(a) shows the relaxation of $\chi^{\prime}(\omega ,t)$ with time $t$ at $T$ = 6.0 K only for $f$ = 0.05 and 0.1 Hz. Figure \ref{fig05}(b) shows the relaxation of $\chi^{\prime\prime}(\omega ,t)$ with time $t$ at $T$ = 6.0 K for $f$ = 0.05 and 0.1 Hz. Both $\chi^{\prime}(\omega ,t)$ and $\chi^{\prime\prime}(\omega ,t)$ decrease with increasing $t$ due to the aging and they are well described by power-law forms 
\begin{equation}
\chi^{\prime}(\omega,t)=\chi_{0}^{\prime}(\omega)+A^{\prime}(\omega)t^{-b^{\prime}} ,
\label{eq03} 
\end{equation}
and
\begin{equation}
\chi^{\prime\prime}(\omega,t)=\chi_{0}^{\prime\prime}(\omega)+A^{\prime\prime}(\omega)t^{-b^{\prime\prime}} ,
\label{eq04} 
\end{equation}
for $t = 0 - 2\times 10^{5}$ sec, respectively. The least-squares fit of the data of $\chi^{\prime}(\omega ,t)$ vs $t$ at $T$ = 6.0 K to Eq.(\ref{eq03}) yields the exponent $b^{\prime}$, and parameters $\chi_{0}^{\prime} (\omega )$ and $A^{\prime}(\omega )$ for each $\omega$, which are listed in Table 1. Figure \ref{fig06} shows the $f$ dependence of the amplitude $A^{\prime}(\omega)$ (listed in Table \ref{Table01}) of the dispersion $\chi^{\prime}(\omega ,t,)$ vs $t$ at $T$ = 6.0 K. We find that the amplitude $A^{\prime}(\omega )$ is well described by a power-law form, 
\begin{equation}
A^{\prime}(\omega)=A_{0}^{\prime}\omega^{-\mu^{\prime}} .
\label{eq05} 
\end{equation}
The least-squares fit of the data of $A^{\prime}(\omega )$ vs $\omega$ for $0.05\le f\le 10$ Hz to Eq.(\ref{eq05}) yields the exponent $\mu^{\prime} = 0.144\pm 0.003$. The exponent $\mu^{\prime}$ is close to the exponent $b^{\prime}$. This result suggests that the time dependence of $A^{\prime}(\omega )t^{-b^{\prime}} $ in Eq.(\ref{eq03}) for $\chi^{\prime}(\omega ,t)$ may be described by a scaling function of $\omega t$ such as $A_{0}^{\prime}(\omega t)^{-b^{\prime}}$. Similarly, the least-squares fit of the data of $\chi^{\prime\prime}(\omega ,t)$ at $T$ = 6.0 K to Eq.(\ref{eq04}) yields the fitting parameters; $b^{\prime\prime}$ = 0.108$\pm $0.013 at $f$ = 0.05 Hz and $b^{\prime\prime}$ = 0.0984$\pm $0.0265 at $f$ = 0.1 Hz. The exponent $b^{\prime\prime}$ is nearly equal to $b^{\prime}$ . Such a small value of $b^{\prime\prime}$ is common to SG systems. For a 3D Ising SG Fe$_{0.5}$Mn$_{0.5}$TiO$_{3}$ ($T_{SG}$ = 20.7 K),\cite{ref05} $b^{\prime\prime}$ is equal to $0.14\pm 0.03$ at 19 K. For a 3D Ising SG Cu$_{0.5}$Co$_{0.5}$Cl$_{2}$-FeCl$_{3}$ graphite bi-intercalation compound ($T_{SG}$ = 3.92 K), the absorption $\chi^{\prime\prime}$ obeys a $\left( \omega t\right)^{-b^{\prime\prime}}$ power-law decay with an exponent $b^{\prime\prime}\approx 0.15 - 0.2$. 

\subsection{\label{resultE}Aging dip in $\chi^{\prime}$ and $\chi^{\prime\prime}$ for the SSW process}

\begin{figure}
\includegraphics[width=7.0cm]{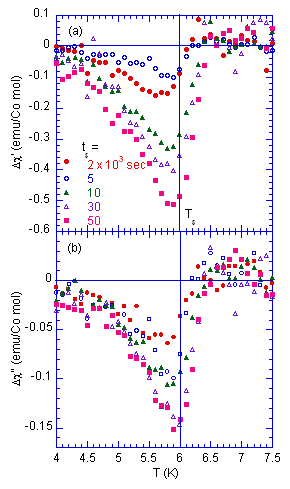}
\caption{\label{fig07}(Color online) $T$ dependence of (a) $\Delta\chi^{\prime}=\chi^{\prime}_{ssw}-\chi^{\prime}_{refh}$ and (b) $\Delta \chi^{\prime\prime}=\chi ^{\prime\prime}_{ssw}-\chi ^{\prime\prime}_{refh}$ for the SSW process at $T_{s}$ (= 6.0 K) for a wait time $t_{s}$ during the ZFC protocol. $t_{s} = 2.0\times 10^{3}$, $5.0\times 10^{3}$, $1.0\times 10^{4}$, $3.0\times 10^{4}$, and $5.0\times 10^{4}$ sec. $f$ = 0.1 Hz. $h$ = 0.1 Oe. $H$ = 0 Oe.}
\end{figure}

\begin{figure}
\includegraphics[width=7.0cm]{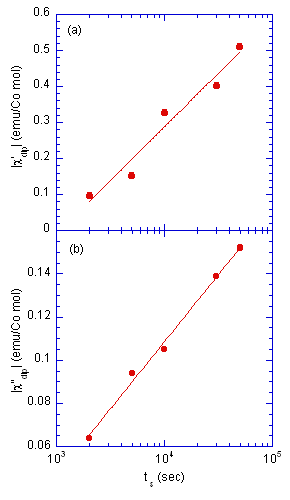}
\caption{\label{fig08}(Color online) Depth of the aging dip at $T$ = 6 K for (a) $\Delta\chi^{\prime}=\chi^{\prime}_{ssw}-\chi^{\prime}_{refh}$ and (b) $\Delta \chi^{\prime\prime}=\chi^{\prime\prime}_{ssw}-\chi ^{\prime\prime}_{refh}$, as a function of the wait time $t_{s}$ for the SSW process ($T_{s}$ = 6.0 K). The solid lines denote least-squares fitting curves of the data to Eq.(\ref{eq07}).}
\end{figure}

Figures \ref{fig07}(a) and (b) show the $T$ dependence of $\Delta\chi^{\prime}=\chi^{\prime}_{ssw}-\chi^{\prime}_{refh}$ and $\Delta \chi^{\prime\prime}=\chi^{\prime\prime}_{ssw} -\chi^{\prime\prime}_{refh}$ for the SSW process, where $h$ = 0.1 Oe, and $H$ = 0 Oe. The wait times are chosen as $t_{s} = 2.0\times 10^{3}$, $5.0\times 10^{3}$, $1.0\times 10^{4}$, $3.0\times 10^{4}$ and $5.0\times 10^{4}$ sec. Both $\Delta \chi^{\prime}$ and $\Delta \chi^{\prime}$ take a local minimum (aging dip) at 5.9 K just below $T_{s}$. Figures \ref{fig08}(a) and (b) show the depth of the aging dip for $\Delta \chi^{\prime}$ and $\Delta \chi^{\prime\prime}$ as a function of $t_{s}$. at $T$ = 5.9 K. We find that the depth is well described by a logarithmic relaxation; 
\begin{equation}
\left| \Delta \chi _{dip} \right| =a_{0} +a_{1} \ln (t),
\label{eq07}
\end{equation}
with $a_{0} = -0.90\pm 0.14$ and $a_{1} = 0.129\pm 0.015$ for $\Delta \chi^{\prime}$ and $a_{0}$ = -0.140$\pm $0.012 and $a_{1} = 0.027\pm 0.001$ for $\Delta \chi^{\prime\prime}$. 

When the system is isothermally aged at $T_{s}$ = 6.0 K for $t_{s}$, it rearranges its spin configuration toward the equilibrium one for this temperature. With further decreasing $T$, the equilibrated state becomes frozen in. The memory is retrieved on reheating. Similar logarithmic behavior is observed in the relaxation of ZFC magnetization with the SSW process during the ZFC protocol for Fe$_{0.5}$Mn$_{0.5}$TiO$_{3}$.\cite{ref06} 

\begin{figure}
\includegraphics[width=7.0cm]{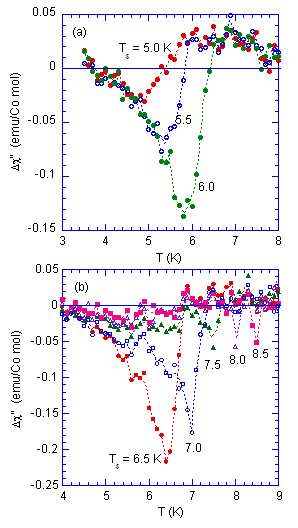}
\caption{\label{fig09}(Color online) $T$ dependence of $\Delta\chi^{\prime\prime}=\chi^{\prime\prime}_{ssw}-\chi^{\prime\prime}_{refh}$ with the SSW process at $T_{s}$ for a wait time $t_{s}$ ($=3.0\times 10^{4}$ sec) during the ZFC protocol. The value of $T_{s}$ is varied as a parameter. (a) $T_{s}$ = 5.0, 5.5, and 6.0 K and (b) $T_{s}$ = 6.5, 7.0, 7.5, 8.0 and 8.5 K. $f$ = 0.1 Hz. $h$ = 0.1 Oe. $H$ = 0 Oe. The dotted lines are guides to the eyes.}
\end{figure}

Figures \ref{fig09}(a) and (b) show the $T$ dependence of $\Delta\chi^{\prime\prime}=\chi^{\prime\prime}_{ssw}-\chi^{\prime\prime}_{refh}$ for the SSW process at $T_{s}$ for a wait time $t_{s}$ ($= 3.0\times 10^{4}$ sec) during the ZFC protocol.The value of $T_{s}$ is changed as a parameter: $T_{s}$ = 5.0, 5.5, 6.0, 6.5, 7.0, 7.5, 8.0 and 8.5 K, where $f$ = 0.1 Hz, $h$ = 0.1 Oe, and $H$ = 0 Oe. The absorption $\Delta\chi^{\prime\prime}$ clearly shows an aging dip. This dip occurs at $T=T_{s}$, where the system has been aged during the SSW process. This result indicates the occurrence of the aging behavior. The depth of the aging dip is the largest at $T_{s}$ = 6.5 K and decreases with further increasing $T_{s}$. The width of the aging dip becomes narrower as the stop temperature $T_{s}$ increases for $7.5\le T_{s}\le 8.5$ K. This implies that the partial memory effect occurs in the intermediate state between $T_{cu}$ and $T_{cl}$. Note that similar behavior is also observed in $\Delta\chi^{\prime}=\chi^{\prime}_{ssw}-\chi^{\prime}_{refh}$, although the data are not shown here. 

\subsection{\label{resultF}Aging dips in $\chi^{\prime}$ and $\chi^{\prime\prime}$ with DSW and TSW processes}

\begin{figure}
\includegraphics[width=7.0cm]{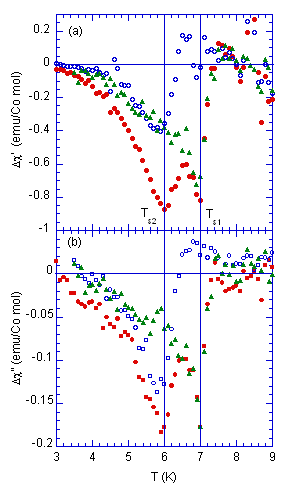}
\caption{\label{fig10}(Color online) $T$ dependence of (a) $\Delta\chi^{\prime}=\chi^{\prime}_{dsw}-\chi^{\prime}_{refh}$ and (b) $\Delta\chi^{\prime\prime}=\chi ^{\prime\prime}_{dsw}-\chi^{\prime\prime}_{refh}$ with the DSW processes, $T_{s1}$ (= 7.0 K) for a wait time $t_{s1}$ ($=3.0\times 10^{4}$ sec) and at $T_{s2}$ (= 6.0 K) for $t_{s2}$ ($=3.0\times 10^{4}$ sec) during the ZFC protocol from 50 to 2.0 K. $f$ = 0.1 Hz. $h$ = 0.1 Oe. $H$ = 0 Oe. $T$ dependence of $\Delta\chi^{\prime}$ and $\Delta\chi^{\prime\prime}$ for the corresponding two SSW process, SSW1 ($T_{s1}$,$t_{s1}$) (closed triangles) and SSW2 ($T_{s2}$,$t_{s2}$) (open circles).}
\end{figure}

We have studied the effect of multiple-stop and wait processes during the ZFC protocol on the AC magnetic susceptibility. Figures \ref{fig10}(a) and (b) show the $T$ dependence of $\Delta \chi^{\prime}=\chi^{\prime}_{dsw} -\chi^{\prime}_{refh} $ and $\Delta \chi^{\prime\prime}=\chi^{\prime\prime}_{dsw} -\chi^{\prime\prime}_{refh} $ for the DSW processes with $T_{s1}$ (= 7.0 K) for a wait time $t_{s1}$ ($=3.0\times 10^{4}$ sec) and at $T_{s2}$ (= 6.0 K) for $t_{s2}$ ($=3.0 \times 10^{4}$ sec) during the ZFC protocol from 50 to 2.0 K, where $f$ = 0.1 Hz, $h$ = 0.1 Oe, and $H$ = 0 Oe. For comparison, we also show the $T$ dependence of $\Delta\chi^{\prime}$ and $\Delta\chi^{\prime\prime}$ for the corresponding two SSW processes: SSW1 with $T_{s1}$ and $t_{s1}$ and SSW2 with $T_{s2}$ and $t_{s2}$. The DSW curves for both $\Delta\chi^{\prime}$ and $\Delta\chi^{\prime\prime}$ show two aging dips at $T_{s1}$ and $T_{s2}$. Each DSW curve is well approximated by a superposition of two corresponding SSW curves which have aging dips at $T_{s1}$ and $T_{s2}$, respectively. The spin configuration denoted by ($t_{s1}$, $T_{s1}$) is imprinted at the first stop $T_{s1}$ and is conserved below $T_{s1}$, even if the system is aged at the second stop $T_{s2}$. The spin configuration denoted by ($t_{s2}$, $T_{s2}$) is also imprinted at the second stop $T_{s2}$ and is also conserved below $T_{s2}$. The spin configuration ($t_{s1}$, $T_{s1}$) is independent of the spin configuration ($t_{s2}$, $T_{s2}$). On reheating from the lowest temperature, the spin configuration ($t_{s2}$, $T_{s2}$) is retrieved at $T$ = $T_{s2}$ and partially rejuvenated above $T_{s2}$. Then the spin configuration ($t_{s1}$, $T_{s1}$) is retrieved at $T$ = $T_{s1}$ and partially rejuvenated above $T_{s1}$. 

\begin{figure}
\includegraphics[width=7.0cm]{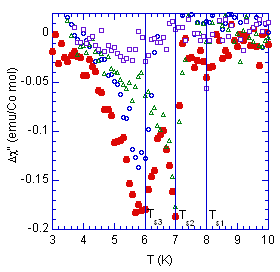}
\caption{\label{fig11}(Color online)$T$ dependence of $\Delta\chi^{\prime\prime}=\chi^{\prime\prime}_{tsw} -\chi^{\prime\prime}_{refh}$ with the TSW (triple-stop and wait) processes.$T_{s1}$ (= 8.0 K) for a wait time $t_{s1}$ ($=3.0\times 10^{4}$ sec), $T_{s2}$ (= 7.0 K) for $t_{s1}$ ($=3.0\times 10^{4}$ sec), and at $T_{s3}$ (= 6.0 K) for $t_{s3}$ ($=3.0\times 10^{4}$ sec) during the ZFC protocol from 50 to 2.0 K. $f$ = 0.1 Hz. $h$ = 0.1 Oe. $H$ = 0 Oe. For comparison, the $T$ dependence of $\Delta\chi^{\prime\prime}=\chi^{\prime\prime}_{ssw} -\chi^{\prime\prime}_{refh}$ with the SSW processes is also shown: $T_{s}$ (= 8.0 K) and $t_{s}$ ($=3.0\times 10^{4}$ sec) (open circles), $T_{s}$ (= 7.0 K) and $t_{s}$ ($=3.0\times 10^{4}$ sec) (open triangles), and $T_{s}$ (= 6.0 K) and $t_{s}$ ($=3.0\times 10^{4}$ sec) (open squares).}
\end{figure}

Figure \ref{fig11} shows the $T$ dependence of $\Delta \chi^{\prime\prime}=\chi^{\prime\prime}_{tsw} -\chi^{\prime\prime}_{refh} $ for the triple-stop and wait (TSW) processes; $T_{s1}$ (= 8.0 K) for a wait time $t_{s1}$ ($=3.0\times 10^{4}$ sec), $T_{s2}$ (= 7.0 K) for a wait time $t_{s1}$ ($=3.0\times 10^{4}$ sec), and at $T_{s3}$ (= 6.0 K) for $t_{s3}$ ($=3.0\times 10^{4}$ sec) during the ZFC cooling protocol from 50 to 2.0 K, where $f$ = 0.1 Hz, $h$ = 0.1 Oe, and $H$ = 0 Oe. The reheating curve $\Delta \chi^{\prime\prime}$ shows three aging dips at $T_{s3}$, $T_{s2}$, and $T_{s1}$, respectively. The depth of the aging dip at $T_{s1}$ is much shallower than those of $T_{s3}$ and $T_{s2}$. 

\subsection{\label{resultG}Nature of aging dip and cumulative aging}

\begin{figure}
\includegraphics[width=7.0cm]{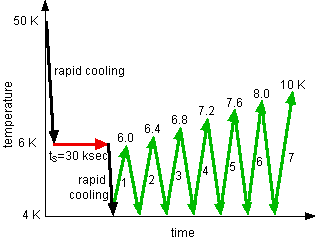}
\caption{\label{fig12}(Color online) Schematic diagram of the procedure for the memory effect in the AC magnetic susceptibility. The system was cooled from 50 K to $T_{s}$ = 6.0 K at $H$ = 0. It was isothermally aged at $T_{s}$ for $t_{s} = 3.0\times 10^{4}$ sec. Then the cooling was resumed from 6.0 to 4.0 K. The data of $\chi^{\prime}$ and $\chi^{\prime\prime}$ were collected repeatedly on reheating from 4.0 K to $T^{*}$ and subsequently on cooling from $T^{*}$ to 4.0 K. After each cycle, $T^{*}$ was increased by 0.4 K for the next cycle: $T^{*}$ = 6.0, 6.4, 6.8, 7.2, 7.6, 8.0, and 10 K.}
\end{figure}

In order to study the nature of the memory and rejuvenation effects in the present system, we have undertaken an extensive measurement of the $T$ dependence of $\chi^{\prime}$ and $\chi^{\prime\prime}$ under the experimental procedure schematically shown in Fig.~\ref{fig12}. The system was cooled from 50 K to $T_{s}$ = 6.0 K at $H$ = 0. It was isothermally aged at $T_{s}$ for $t_{s} = 3.0\times 10^{4}$ sec (the SSW process). Then the cooling was resumed from 6 to 4.0 K. The data of $\chi^{\prime}$ and $\chi^{\prime\prime}$ were collected repeatedly on heating from 4.0 K to $T^{*}$ and subsequently on cooling from $T^{*}$ to 4.0 K. After each cycle, $T^{*}$ was increased by 0.4 K for the next cycle: $T^{*}$ = 6.0, 6.4, 6.8, 7.2, 7.6, and 8.0 K. Finally the data of $\chi^{\prime}$ and $\chi^{\prime\prime}$ were taken with increasing $T$ from 4.0 to 10.0 K after cooling the system from $T^{*}$ = 8.0 to 4.0 K. The cooling rate of the system during the measurement of $\chi^{\prime}$ and $\chi^{\prime\prime}$ is the same as the heating rate; typically 1.3 K/h. 

Before our measurements, the following two things are expected. (i) The aging dip appears in $\Delta\chi^{\prime\prime}$ at $T_{s}$ (= 6.0 K) in the $T^{*}$ (= 6.0, 6.4, 6.8 K)-scan (the memory effect). According to the droplet model of SG's,\cite{ref25,ref26} the domains are unaffected by small temperature shift (at least $\Delta T=T^{*}-T_{s}=0.8$ K) if the overlap length ($L_{\Delta T}$) is larger than the size of domains that can respond to the AC field of angular frequency $\omega$. The overlap length depends on the temperature difference $\Delta T$ as $L_{\Delta T}\approx 1/(\Delta T)$.\cite{ref27} (ii) The aging dip disappears in $\Delta\chi^{\prime\prime}$ at $T_{s}$ (= 6.0 K) in the $T_{s}^{*}$ ($> 7.2$ K)-scan (the rejuvenation effect), because the memory imprinted at $T_{s}$ may be erased since the system is kept at $T$ well above $T_{s}$ before cooling the system down to 4.0 K. 

Contrary to our expectation, our reseult is rather different from our prediction. While performing these experiments, we observe a surprising behavior of the AC magnetic susceptibility, which may be related to the nature of the cumulative aging in this system. What is a typical definition of the cumulative aging? 
Suppose that the system is aged at $T = T_{s}$ for a wait time $t_{s}$ after the ZFC protocol. The age of the system at $T_{s}$ is $t_{s}$. For typical SG's, the age of the system at $T = T_{s}$ remains unchanged when the temperature shifts from $T_{s}$ to the lower temperature $T_{0}$ (the negative $T$-shift). For the cumulative aging, in contrast the age of the system at $T = T_{s}$ increases with increasing $t$ so long as $T$ is kept at $T_{0}$ below $T_{s}$, due to the effect of relaxations at $T_{0}$.\cite{ref26} 

\begin{figure}
\includegraphics[width=7.0cm]{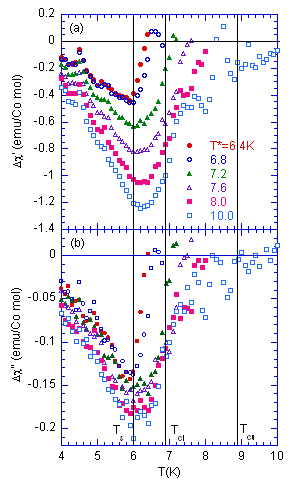}
\caption{\label{fig13}(Color online)$T$ dependence of $\Delta \chi^{\prime}=\chi^{\prime}_{ssw}-\chi^{\prime}_{refh}$ and $\Delta \chi^{\prime\prime}=\chi^{\prime\prime}_{ssw} -\chi^{\prime\prime}_{refh}$, recorded on reheating from 4 K to $T^{*}$. $T^{*}$ = 6.0, 6.4, 6.8, 7.2, 7.6, 8.0, and 10.0 K. See the successive procedures illustrated in Fig.~\ref{fig12}. Reference curve: ZFC protocol (quenching from 50 to 2.0 K) at $H$ = 0 and measurements of $\chi^{\prime}_{refh}$ and $\chi^{\prime\prime}_{refh}$ with increasing $T$ from 2.0 to 12.0 K at $H$ = 0.}
\end{figure}

Figures \ref{fig13}(a) and (b) show the $T$ dependence of $\Delta\chi^{\prime}$ and $\Delta \chi^{\prime\prime}$ following the experimental procedures illustrated by Fig.~\ref{fig12}. For simplicity, we only show the data taken with increasing $T$ from 4.0 K to $T^{*}$ (denoted by $T^{*}$-scan) where $T^{*}$ = 6.0, 6.4, 6.8, 7.2, 7.6, 8.0, and 10.0 K. The overall $T$ dependence of $\Delta\chi^{\prime\prime}$ is similar to that of $\Delta\chi^{\prime}$, although the magnitude of $\Delta\chi^{\prime\prime}$ is much smaller than $\Delta\chi^{\prime}$. Since the data of $\Delta\chi^{\prime}$ vs $T$ is noisy, we discuss only the data of $\Delta\chi^{\prime}$ vs $T$ in the $T^{*}$-scan. The difference $\Delta\chi^{\prime}$ shows an aging dip at $T_{s}$ (= 6 K) in $T^{*}$-scan where $T^{*}$ = 6.0 - 10.0 K. As shown in Fig.~\ref{fig13}(a), the depth of the aging dip increases as $T^{*}$ increases. The width of the depth in the aging dip becomes broader and broader as $T^{*}$ increases. The full width at the half maximum in the depth of the aging dip is on the order of $\Delta T = 2.4$ K for $T^{*}$ (= 10.0 K)-scan. We also note that the temperature at which the depth of the aging dip takes maximum, slightly shifts to the high-$T$ side with increasing $T^{*}$. This means that once the memory of the spin configuration is imprinted at $T_{s}$ during the ZFC protocol with the SSW process, it cannot be erased as long as $T^{*}$ is kept below $T_{cu}$. 

The magnitude of $\Delta\chi^{\prime}$ between $T_{cl}$ and $T_{cu}$ also drastically increases with increasing $T^{*}$ for $7.2\le T^{*}\le 10.0$ K. This also implies that the magnitude of $\Delta\chi^{\prime}$ between $T_{cl}$ and $T_{cu}$ increases with increasing the total time for which the system is spent at $T$ between $T_{cl}$ and $T_{cu}$. Such an increase in $\left| \Delta\chi^{\prime}\right|$ with $t$ indicates that the cumulative aging continues to occur in this system. At the present stage, we cannot explain sufficiently the complicated aging behavior of this system. However, we can only say that (i) the aging dip appears due to the memory effect in the SG phase below $T_{cl}$ and that (ii) the cumulative aging occurs mainly between $T_{cl}$ and $T_{cu}$ due to the possible domain growth process. This process is temperature cumulative, in the sense that aging continues additively from one temperature to the other.

\section{\label{dis}DISCUSSION}
In stage-2 CoCl$_{2}$ GIC, there are two graphite layers between adjacent CoCl$_{2}$ layers. Each CoCl$_{2}$ layer is formed of small ferromagnetic domains, which are randomly distributed and are magnetically frustrated.\cite{ref28} Our results on the aging and memory obtained above for the stage-2 CoCl$_{2}$ GIC are summarized as follows. (i) The nature of the aging and memory of the intermediate phase between $T_{cu}$ and $T_{cl}$ is characterized by a mainly cumulative aging and a positive hysteresis of both $\Delta \chi^{\prime}_{refch}$ and $\Delta \chi^{\prime\prime}_{refch}$. The partial memory effect is observed as a weak aging dip in the immediate vicinity of $T_{s}$. The growth of domains gives rise to a cumulative aging, any reduction in $\chi^{\prime}$ ($=\chi^{\prime}-\chi_{refh}$) and $\chi^{\prime\prime}$ ($=\chi^{\prime\prime}-\chi_{refh}$) remains as long as $T$ is kept below $T_{cu}$. The domain size serves as a simple order parameter. The sign of both $\Delta\chi^{\prime}_{refch}$ and $\Delta \chi^{\prime\prime}_{refch}$ changes from positive to negative at $T=T_{cl}$ with decreasing $T$. The $T$ dependence of both $\Delta \chi^{\prime}_{refch}$ and $\Delta\chi^{\prime\prime}_{refch}$ is similar to that of the nonlinear AC magnetic susceptibility $\chi_{3}^{\prime}$ .\cite{ref21,ref22} (ii) The nature of the aging and memory of the low temperature SG phase below $T_{cl}$ is characterized by a prominent aging dip, a negative sign of $\Delta \chi^{\prime}_{refch} $ and $\Delta \chi^{\prime\prime}_{refch}$, and a $\omega t$-scaling of of $\chi^{\prime}$ and $\chi^{\prime\prime}$ . Aging a SG at a temperature below $T_{cl}$ reduces the AC susceptibility only in the immediate vicinity of $T_{s}$, creating an aging dip. As long as $T$ is kept below $T_{s}$, there will be memory of this dip (the full memory effect). This SG phase also approximately exhibits $\omega t$ scaling, where the aging of $\chi^{\prime\prime}(\omega ,t)$ is a function of the product $\omega t$ and not $\omega$ and waiting time $t$, separately. 

The aging and memory below $T_{cl}$ and the cumulative aging between $T_{cu}$ and $T_{cl}$ in stage-2 CoCl$_{2}$ GIC are reminiscent of that seen in reentrant ferromagnet CdCr$_{1.9}$In$_{0.1}$S$_{4}$.\cite{ref11} This system undergoes two transitions at $T_{RSG}$ = 10 K and $T_{c}$ = 70 K, forming the reentrant spin glass (RSG) phase below $T_{RSG}$ and the ferromagnetic (FM) phase between $T_{RSG}$ and $T_{c}$. In the FM phase, $\chi^{\prime\prime}_{refc} $ is relatively larger than $\chi^{\prime\prime}_{refh} $ , leading to the positive sign of $\Delta \chi^{\prime\prime}_{refch} $ . The aging dip is seen around a stop temperature $T_{s}$ below $T_{RSG}$, reflecting the aging and memory effect. In contrast, no aging dip is observed at $T_{s}$ between $T_{c}$ and $T_{RSG}$, in the case when the lowest starting temperature $T_{0}$ (= 30 K) on reheating is well lower than $T_{s}$, indicating that no memory effect exists. However, if the $T_{0}$ reaches closer to $T_{s}$ (= 67 K), an aging dip is observed around $T$ = $T_{s}$. The enhancement of the partial memory effect occurs as $T_{0}$ becomes closer and closer to $T_{s}$. In the RSG phase below 10 K, the full memory effect is observed as in usual SG's. For a reentrant ferromagnet stage-2 Cu$_{0.8}$Co$_{0.2}$Cl$_{2}$ GIC,\cite{ref13} in contrast, the aging dip is observed in the both the RSG phase and the FM phase in the TRM measurement after the SSW process. The full memory effects are observed in both the RSG and FM phases. 

It is important to stress that the main features of cumulative relaxation at high temperatures and the aging and memory at lower temperatures in the reentrant ferromagnets are common to systems other than SG systems. Similar aging and memory has been reported in the imaginary part of complex dielectric susceptibility for the relaxor ferroelectrics (PbMn$_{1/3}$Nb$_{2/3}$O$_{3}$)$_{1-x}$(PbTiO$_{3}$)$_{x}$, with $x$ = 0.1 and 0.12 by Chao et al.\cite{ref16} There occurs a crossover from the cumulative aging regime at higher temperatures to the aging and memory regime at lower temperatures. The positive sign of $\Delta \epsilon^{\prime\prime}_{refch}$ is seen in the cumulative aging regime, leading to the hysteresis between the reference cooling and reference heating curve of $\epsilon^{\prime\prime}_{refc}$ and $\epsilon^{\prime\prime}_{refh}$. In this sense, the aging and memory effect obtained in the present work is universal to other disordered systems. 

\section{CONCLUSION}
Memory and aging behavior in stage-2 CoCl$_{2}$ GIC ($T_{cu}$ = 8.9 K and $T_{cl}$ = 6.9 K) has been studied by AC magnetic susceptibility as well as TRM magnetization. There occurs a crossover from a cumulative aging (and a partial memory effect) for the domain-growth ferromagnetic regime in the intermediate state between $T_{cu}$ and $T_{cl}$, to an aging and memory in a SG regime below $T_{cl}$. The difference between the reference cooling and reference heating curves, $\Delta\chi^{\prime\prime}_{refch}$, shows a temperature-dependent hysteresis. The sign of $\Delta \chi^{\prime\prime}_{refch} $ changes from positive to negative on crossing $T_{cl}$ from the high $T$-side. The time dependence of $\chi^{\prime}$ and $\chi^{\prime\prime}$ below $T_{cl}$ is described by a scaling function of $\omega t$. The aging and memory thus obtained are common to those observed in reentrant ferromagnets. 

\begin{acknowledgments}
We are grateful to Prof. H. Suematsu for providing us with single crystals of kish graphites and to Prof. M. Matsuura for valuable discussion.
\end{acknowledgments}

\end{document}